\documentclass{PoS}

\usepackage[utf8]{inputenc}
\usepackage{amsmath,amsfonts,amssymb,mathrsfs,fixmath}

\usepackage[sort&compress]{natbib}
\bibpunct{[}{]}{,}{n}{}{}
\newcommand{\cpp}{\ensuremath{\mu_A^{(0)}}}
\newcommand{\cp}{\ensuremath{\mu_A}}
\newcommand{\mr}{\ensuremath{m_r}}
\newcommand{\mi}{\ensuremath{m_i}}

\newcommand{\comment}[1]{}
\newcommand{\lr}[1]{ \left( #1 \right) }
\newcommand{\lrs}[1]{ \left[ #1 \right] }
\newcommand{\lrc}[1]{ \left\{ #1 \right\} }
\newcommand{\vev}[1]{ \langle \, #1 \, \rangle }

\newcommand{\tr}{ {\rm Tr} \, }

\newcommand{\expa}[1]{ \exp{\left( #1 \right)} }

\title{A lattice mean-field study of the phase diagram of interacting parity-breaking Weyl semimetals.}
\ShortTitle{Phase diagram of interacting $\mathcal{P}$-breaking Weyl semimetals}

\author{\speaker{Pavel Buividovich}\thanks{This work is supported by the S.~Kowalevskaja award from the Alexander von Humboldt foundation.}\\
       Institut f\"{u}r Theoretische Physik, Universit\"{a}t Regensburg\\
       E-mail: \email{pavel.buividovich@physik.uni-regensburg.de}}

\author{Matthias Puhr\\
        Institut f\"{u}r Theoretische Physik, Universit\"{a}t Regensburg \\
        E-mail: \email{matthias.puhr@physik.uni-regensburg.de}}

\abstract{We perform a mean-field study of the phase diagram of interacting Weyl semimetals with broken parity, that is, with different densities of right- and left-handed quasiparticles. As a simple model system, we consider the Wilson-Dirac Hamiltonian with the chiral chemical potential and on-site repulsive interactions. We find that the chiral chemical potential somewhat shrinks the region of the pion condensation (Aoki phase) in the parameter space of the bare mass and the interaction strength, so that the condensation thresholds are at smaller interaction strengths. The renormalized chiral chemical potential monotonously grows with interaction strength everywhere in the phase diagram, and only the growth rate is discontinuous across the phase transition lines. These findings are in full agreement with previous results obtained by one of the authors for the continuum Dirac Hamiltonian, except for the fact that for our lattice model with explicitly broken chiral symmetry the boundaries of the Aoki phase remain sharp second-order phase transitions even at nonzero chiral chemical potential and there are no signatures of Cooper-type instabilities in the weakly interacting regime.}

\FullConference{The 32nd International Symposium on Lattice Field Theory,\\
		23-28 June, 2014\\
		Columbia University New York, NY}

\graphicspath{{./plots/}}

\begin{document}

\section{Introduction}
\label{sec:intro}

 Anomalous transport phenomena in systems of chiral fermions have recently become a subject of intense research. Examples of anomalous transport include the Chiral Magnetic, the Chiral Separation and the Chiral Vortical Effects (abbreviated as CME, CSE and CVE). While the role of these phenomena in the collective flow in off-central heavy-ion collisions \cite{Kharzeev:08:2} is still disputable \cite{Skokov:13:1,Hongo:13:1}, they could also be realized in table-top experiments with Weyl semimetals \cite{Wan:11:1,Burkov:11:1,Vazifeh:13:1,Landsteiner:13:1,Basar:13:1,Chernodub:13:1}, recently discovered materials in which quasiparticles behave as relativistic Weyl fermions. In particular, the CME can be realized in Weyl semimetals with the broken parity, in which the numbers of right- and left-handed quasiparticles are different. This difference can be conveniently parameterized in terms of the chiral chemical potential $\mu_A$ \cite{Kharzeev:08:2}. A natural way to achieve such chirality imbalance is by placing the sample in parallel electric and magnetic fields, or by applying external voltage or strain to a multilayer of ordinary and topological insulators \cite{Balents:12:1}.

 An important feature of anomalous transport coefficients is that under the conditions of the validity of the hydrodynamical or the Fermi liquid approximations they take universal values even in strongly coupled systems \cite{Son:09:1,Sadofyev:10:1,Banerjee:12:1,Jensen:12:1}. However, spontaneous chiral symmetry breaking violates these approximations \cite{Buividovich:13:8,Buividovich:14:1} due to the emergence of massless Goldstone modes. Moreover, anomalous transport coefficients are subject to perturbative renormalization if the corresponding currents (e.g. the electric current in the case of CME) are coupled to dynamical gauge fields \cite{Miransky:13:1,Jensen:13:1,Gursoy:14:1,Buividovich:14:1}. Both sources of corrections to anomalous transport coefficients might be relevant for Weyl semimetals. First of all, it is obvious that only electromagnetic interactions are relevant in condensed matter systems, thus one cannot neglect the coupling of dynamical photons to electric current and charge density. Second, electrostatic interactions in condensed matter systems are effectively enhanced by a factor of the inverse Fermi velocity $v_F^{-1} \gg 1$, which might lead to spontaneous chiral symmetry breaking \cite{Son:07:1}.

 A mean-field study of the chiral magnetic effect for the continuum Dirac Hamiltonian with contact interactions between electric charges was recently performed in \cite{Buividovich:14:1}. It was found that interactions increase the chiral chemical potential as well as the chiral magnetic conductivity both in the weak-coupling regime and in the strongly coupled phase with broken chiral symmetry. Moreover, it turned out that at nonzero chiral chemical potential the second-order phase transition associated with spontaneous chiral symmetry breaking turns into a soft crossover, and the chiral condensate starts growing with interaction potential even at arbitrarily weak interactions. Such picture is in fact typical for Cooper-type instability in the presence of particle-like and hole-like Fermi surfaces (which are the Fermi surfaces of the left- and right-handed fermions in our case).

 However, the calculations of \cite{Buividovich:14:1} were done in the Dirac cone approximation with an artificial cutoff scale $\Lambda$. In real crystals the dispersion relation always deviates from the linear one by virtue of the compactness of the Brillouin zone. As a result, the chiral symmetry is manifestly broken at higher energies. Therefore it is important to take into account the nonlinearity of the dispersion relation and the absence of exact chiral symmetry when considering the effect of interactions on real Weyl semimetals.

 In these Proceedings we perform a mean-field study of the phase diagram of a simple lattice model of Weyl semimetals with chiral imbalance and on-site repulsive interactions between electric charges, leaving the calculation of the chiral magnetic conductivity for future work. We note that while the phase diagram of Weyl semimetals with momentum separation between Weyl nodes has been studied in detail \cite{Wang:13:1, Wei:12:1, Sekine:13:1}, the effects of interactions in Weyl semimetals with energy separation between Weyl nodes have not been considered before \cite{Buividovich:14:1}. As the simplest model of parity-breaking Weyl semimetals, we consider the Wilson-Dirac Hamiltonian \nocite{Sekine:13:1, Vazifeh:13:1, Hosur:13:1} with chiral chemical potential. Since for this Hamiltonian the chiral symmetry is explicitly broken at high energies by the Wilson term, the relevant phase transition is associated with the pion condensation (Aoki phase) rather than spontaneous chiral symmetry breaking \cite{Aoki:84:1}. In agreement with the results obtained for the continuum Dirac Hamiltonian \cite{Buividovich:14:1}, we find that the chiral chemical potential is enhanced by interactions everywhere in the parameter space of the model, and that the Aoki phase is shifted to weaker interactions in the presence of chiral imbalance. In contrast to the continuum case, the transition to the Aoki phase remains a sharp second-order phase transition.

 Let us also note that at least within the kinetic theory approximation chirally imbalanced matter appears to be unstable towards the formation of magnetic fields with nontrivial magnetic helicity \cite{Yamamoto:13:1,Sadofyev:13:1}. Since the coupling to magnetic fields is suppressed by the smallness of the Fermi velocity, one can expect that for condensed matter systems this instability will develop rather slowly. One can also imagine a situation in which the chirality is pumped into the system at a constant rate which compensates for its decay. We thus assume that the decay of chirality imbalance is negligible and consider an approximate equilibrium state of the system with constant chiral chemical potential.

\section{Mean-field approximation for Wilson-Dirac Hamiltonian with on-site interactions}
\label{sec:mf}

 We consider the following many-body Hamiltonian with on-site interactions:
\begin{equation}
\label{ManyBodyHamiltonian}
 \hat{H}
 =
 \sum\limits_{x,y} \hat{\psi}^{\dag}_x h^{\lr{0}}_{x,y} \hat{\psi}_y
 +
 V \sum\limits_x \lr{\hat{\psi}^{\dag}_x \hat{\psi}_x - 2}^2 ,
\end{equation}
where $\hat{\psi}_x = \lrc{\hat{\psi}_{\uparrow R,x}, \hat{\psi}_{\downarrow R,x}, \hat{\psi}_{\uparrow L,x}, \hat{\psi}_{\downarrow L,x} }$ are the Dirac spinor-valued fermionic annihilation operators, $V > 0$ is the on-site repulsive interaction potential and $\lr{\hat{\psi}^{\dag}_x \hat{\psi}_x - 2} \equiv \hat{q}_x$ is the operator of charge at site $x$ \footnote{where the summand $-2$ is due to the charge of the ions which form the crystal}. $h^{\lr{0}}_{x,y} \equiv h^{\lr{0}}_{x-y}$ is the single-particle Wilson-Dirac Hamiltonian, which in the momentum space reads
\begin{equation}
\label{WilsonDiracHamiltonian}
 h^{\lr{0}}_{x,y} = \sum\limits_k \, e^{i k \lr{x - y}} h^{\lr{0}}\lr{k},
 \quad
 h^{\lr{0}}\lr{k} =
 \sum\limits_{i=1}^3 \lr{v_F \alpha_i \sin\lr{k_i} + 2 \gamma_0 \sin^2\lr{k_i/2} }
 + \mu_A^{\lr{0}} \gamma_5 + m^{\lr{0}} \gamma_0 ,
\end{equation}
where $v_F$ is the Fermi velocity, $\alpha_i = -i \gamma_0 \gamma_i$, $\gamma_{\mu}$ and $\gamma_5$ are the Euclidean gamma-matrices in the chiral representation, $k_i \in \lrs{-\pi, \pi}$ are the spatial lattice momenta, $\mu_A^{\lr{0}}$ is the bare chiral chemical potential and $m^{\lr{0}}$ is the bare mass. Since the Fermi velocity $v_F$ can be removed from (\ref{ManyBodyHamiltonian}) and (\ref{WilsonDiracHamiltonian}) by rescaling $V$, $m^{\lr{0}}$ and $\mu_A^{\lr{0}}$ \cite{Buividovich:14:1}, we set it to unity in what follows.

 In order to arrive at the mean-field approximation, we perform the Suzuki-Trotter decomposition of the partition function $\mathcal{Z} = \tr\expa{-\hat{H}/T}$, followed by the Hubbard-Stratonovich transformation in the particle-hole channel. The integral over the Hubbard-Stratonovich field $\Phi_{x, \alpha\beta} = \bar{\Phi}_{x, \beta\alpha}$ (where $\alpha$, $\beta$ are the spinor indices) is then replaced by its saddle-point approximation, as in \cite{Buividovich:14:1}. Taking the limit of zero temperature and assuming that at the saddle point $\Phi_{x, \alpha\beta} \equiv \Phi_{\alpha\beta}$ is homogeneous both in space and in time, we find that the saddle point values of $\Phi_{\alpha\beta}$ can be found from the minimum of the following functional:
\begin{eqnarray}
\label{MeanFieldFunctional}
 \mathcal{F}\lrs{\Phi} = \frac{1}{L_s^3} \, \sum\limits_{\epsilon_i < 0} \epsilon_i + \frac{\Phi_{\alpha\beta} \Phi_{\beta\alpha}}{4 V} ,
\end{eqnarray}
where we assume summation over repeated indices and the sum in the first term goes over all negative energy levels of the effective single-particle Hamiltonian $h\lr{k} = h^{\lr{0}}\lr{k} + \Phi$. This sum is nothing but the energy of the Dirac sea, which is finite in any realistic lattice model. We further assume that rotational symmetry of $h\lr{k}$is not broken, which restricts the saddle-point values of $\Phi$ to have the form $\Phi = \lr{\mr - m^{\lr{0}}} \gamma_0 + \mi \gamma_0 \gamma_5 + (\cp - \cpp)\gamma_5$, where $\mr$ is the renormalized mass, $\mu_A$ is the renormalized chiral chemical potential and $m_i$ is the parity-breaking mass term which corresponds to the nonzero pion condensate $\vev{\hat{\psi}^{\dag} \gamma_0 \gamma_5 \hat{\psi}}$.

 The energy levels of the effective single-particle Hamiltonian $h\lr{k}$ with such saddle-point value of $\Phi$ are $\epsilon_{s,\sigma}\lr{\vec{k}} = s \sqrt{ \lr{S - \sigma \mu_A}^2 + m_i^2 + \lr{m_r + W}^2  }$, where $s = \pm 1$, $\sigma = \pm 1$, $S = \sqrt{\sum\limits_i \sin^2\lr{k_i} }$ and $W = \sum\limits_i 2 \sin^2\lr{k_i/2}$. Let us note that while for the continuum Dirac Hamiltonian the parity-breaking mass term $m_i$ can be rotated to the standard mass term $m_r \gamma_0$ by a chiral rotation, this is no longer the case for Wilson-Dirac fermions with explicitly broken chiral symmetry. The residual of the $U\lr{1}_A$ chiral symmetry is only the $Z_2$ discrete symmetry $m_i \rightarrow -m_i$, which is spontaneously broken in the Aoki phase \cite{Aoki:84:1}. At the boundary of the Aoki phase the quadratic term in the effective potential for $m_i$ vanishes, and the fluctuations of $m_i$ become massless. This is how the massless pion is realized for lattice fermions with no explicit chiral symmetry: the pion should no longer be interpreted as a Goldstone mode which is massless everywhere in a phase with spontaneously broken symmetry, but rather as fluctuations of the order parameter which become long-range in the vicinity of the phase transition. On the other hand, due to explicitly broken chiral symmetry the conventional mass term $m_r$ is subject to strong additive renormalization, and massless excitations can only be realized if the bare mass term $m^{\lr{0}}$ is negative.

\section{Numerical results}
\label{sec:results}

\begin{figure*}[h!tpb!]
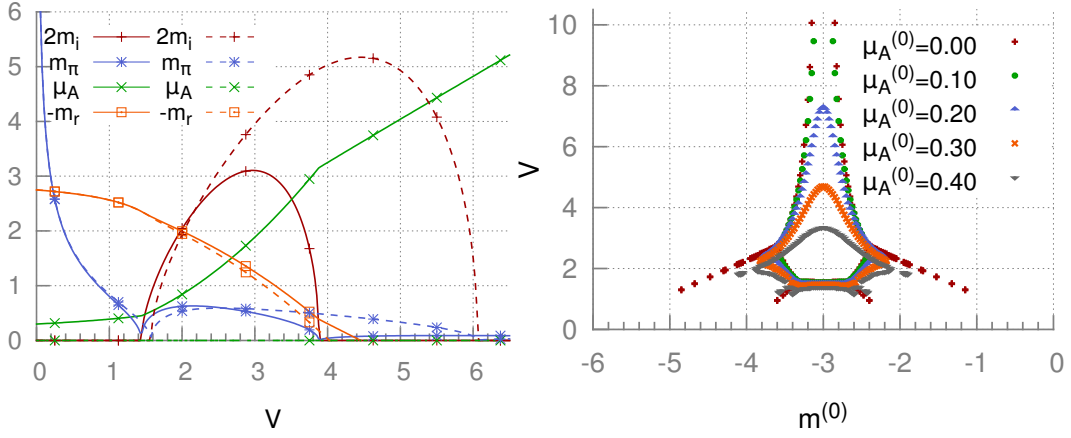

  \centering
  \includegraphics[width=6.8cm]{{{pion_Mb=-2.75_muA=0.30_}}}
  \includegraphics[width=7.2cm]{{{phdiag_comp}}}\\
  \label{fig:phdiag_comp}
  \caption{On the left: the renormalized mass $m_r$, the renormalized chiral chemical potential $\mu_A$, the parity-breaking mass term $m_i$ and the pion mass $m_{\pi}$ as functions of the interaction potential $V$ at $m^{\lr{0}} = -2.75$ and $\mu^{\lr{0}}_A = 0$ (dashed lines) and $\mu^{\lr{0}}_A = 0.30$ (solid lines). The relative scale of different quantities is changed in order to make the plot more illustrative. On the right: phase diagram of the model (1) in the parameter space of the bare mass $m^{\lr{0}}$ and inter-electron interaction potential $V$ for different values of the bare chiral chemical potential $\cpp$. The points mark the border between the phase with $\mi=0$ and the Aoki phase with $\mi \neq 0$.}
\end{figure*}

 In order to find the saddle-point values of the parameters $\mu_A$, $m_r$ and $m_i$ of the effective single-particle Hamiltonian $h\lr{k}$, we perform numerical minimization of the functional (\ref{MeanFieldFunctional}) using the differential evolution method \cite{Storn:97:1}. We consider only the negative values of $m^{\lr{0}}$, since nontrivial phases with massless excitations only exist in this case. The phase structure of the model (\ref{ManyBodyHamiltonian}) in the parameter space of the bare mass $m^{\lr{0}}$ and the interaction potential $V$ is illustrated on Fig.~\ref{fig:phdiag_comp} for different values of the bare chiral chemical potential $\mu_A^{\lr{0}}$.

 On the left plot we show the pion mass $m_{\pi}^2 = \frac{\partial^2 \mathcal{Z}}{\partial m_i^2}$ and the parameters $\mu_A$, $m_r$ and $m_i$ of the effective Hamiltonian $h\lr{k}$ as functions of $V$ at  $m^{\lr{0}} = -2.75$ and $\mu_A^{\lr{0}} = 0$ and $\mu_A^{\lr{0}} = 0.30$. Both for zero and nonzero $\mu_A^{\lr{0}}$ we see a sharp second-order phase transition to the Aoki phase with $m_i \neq 0$. The renormalized mass $m_r$ behaves monotonously in both phases. As expected, the pion mass goes to zero at the boundaries of the Aoki phase. At nonzero $\mu_A^{\lr{0}}$ the Aoki phase shrinks and its boundaries are shifted towards smaller $V$.

\begin{figure*}[h!tpb!]
 \centering
 \includegraphics[width=7cm]{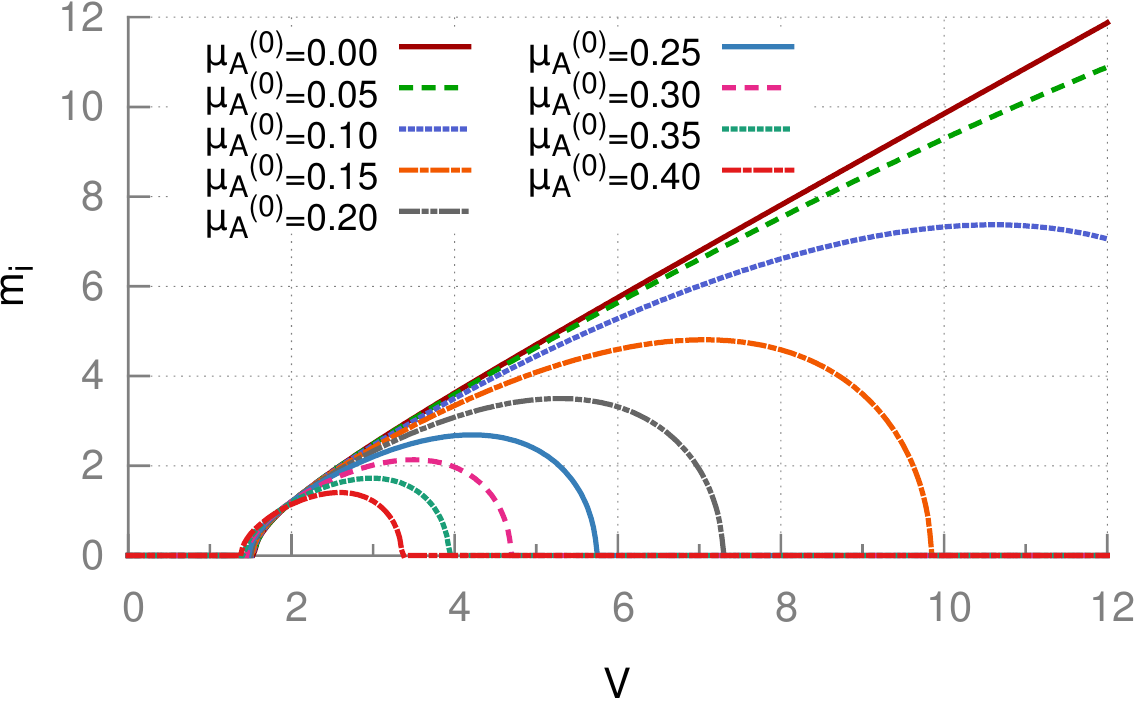}
 \includegraphics[width=7cm]{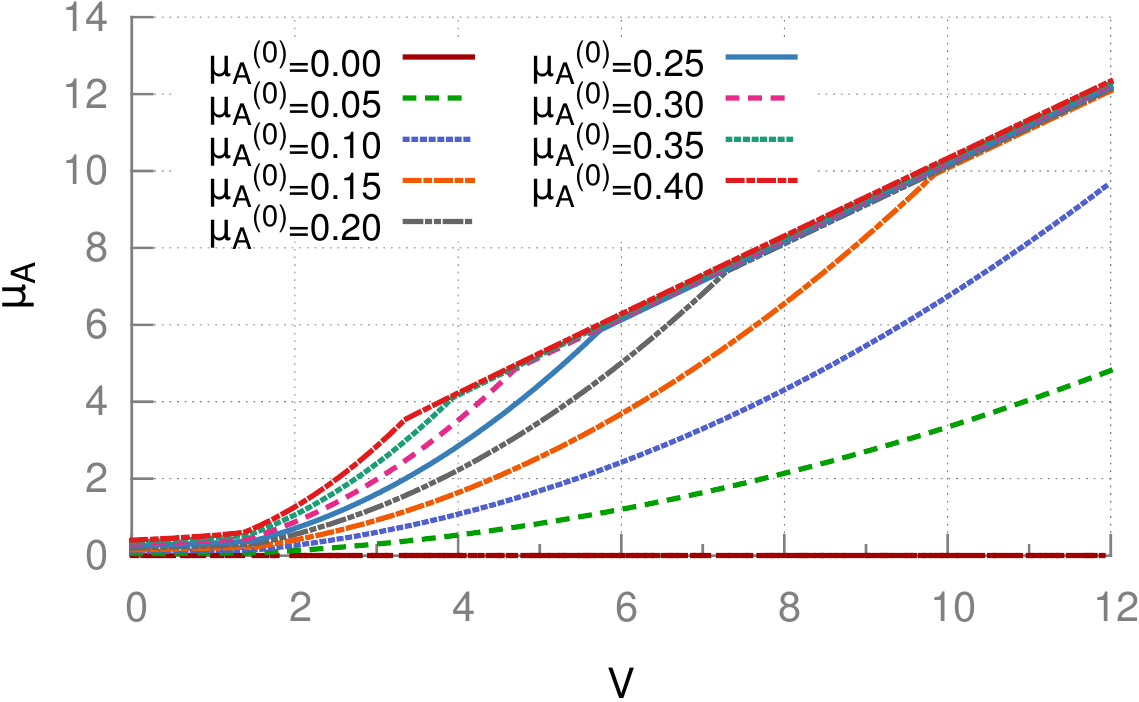}\\
 \label{fig:mi_muA}
 \caption{The parity-breaking mass term $m_i$ (on the left) and the renormalized chiral chemical potential $\mu_A$ (on the right) as functions of the inter-electron interaction potential $V$ at the bare mass $m^{\lr{0}} = -3$.}
\end{figure*}

 The points on the right plot mark the boundaries of the Aoki phase region with $m_i \neq 0$ in the space of $m^{\lr{0}}$ and $V$. One can see the characteristic ``Aoki fingers'' touching the $V = 0$ axis at the usual critical values of $m^{\lr{0}} = 0, \, -2, \, -4, \, -6$. These ``fingers'' are, however, very thin and difficult to distinguish in numerical minimization. For this reason we were not able to follow them all the way down to $V = 0$. Nonzero chiral chemical potential tends to shrink the Aoki phase region and shift it towards smaller $V$ for all values of $m^{\lr{0}}$. To illustrate this conclusion, on Fig.~\ref{fig:mi_muA} on the left we plot $m_i$ as a function of $V$ for several nonzero values of $\mu_A^{\lr{0}}$.

 An important observation is that the renormalized chiral chemical potential $\mu_A$ is always increased by interactions, as illustrated on Fig.~\ref{fig:mi_muA} on the right. Moreover, it seems that in the strong-coupling regime the function $\mu_A\lr{V}$ always approaches some universal linear asymptotics which is almost independent of the bare value $\mu_A^{\lr{0}} \equiv \mu_A\lr{0}$.

\section{Discussion and conclusions}
\label{sec:conclusions}

 In agreement with the results of \cite{Buividovich:14:1} obtained for the continuous Dirac Hamiltonian, we have found that also for lattice fermions with only approximate low-energy chiral symmetry the interactions tend to increase the chiral chemical potential term. Such renormalization of $\mu_A$ is not surprising, since it is not the conventional chemical potential coupled to the conserved charge, and there are no Ward identities which would protect it from renormalization. At the same time, nonzero $\mu_A$ lowers the vacuum energy and is thus energetically favourable \cite{Buividovich:14:1}. It seems though that the renormalization of $\mu_A$ is multiplicative, as for zero bare value $\mu_A^{\lr{0}}$ the renormalized $\mu_A$ remains exactly zero in the presence of interactions. At nonzero $\mu_A^{\lr{0}}$ the boundaries of the Aoki phase with condensed pion field are shifted to smaller $V$, also in qualitative agreement with \cite{Buividovich:14:1}.

 However, in contrast to the continuum case, the phase transition to the Aoki phase (which in the continuum is equivalent to the phase with broken chiral symmetry) remains a sharp second-order phase transition at nonzero $\mu_A^{\lr{0}}$. Possible explanation of this difference is that in the continuum case the finite densities of left-handed particles and right-handed holes (or vice versa) at nonzero $\mu_A$ trigger Cooper-type instability towards the formation of the chiral condensate $\Sigma = \vev{\hat{\psi}^{\dag}_R \hat{\psi}_L} + \vev{\hat{\psi}^{\dag}_L \hat{\psi}_R} $ at arbitrarily weak interactions. In the lattice Hamiltonian (\ref{WilsonDiracHamiltonian}) the chiral symmetry is already broken, thus the Cooper instability has little effect on the effective mass $m_r$. At the same time, the pion condensate $\vev{\pi^{0}} = \vev{\hat{\psi}^{\dag} \gamma_0 \gamma_5 \hat{\psi}} = \vev{\hat{\psi}^{\dag}_R \hat{\psi}_L} - \vev{\hat{\psi}^{\dag}_L \hat{\psi}_R}$ is also insensitive to the existence of particle- and hole-like Fermi surface since there are equal numbers of left-handed and right-handed particles and holes in a system and the two terms in the difference $\vev{\hat{\psi}^{\dag}_R \hat{\psi}_L} - \vev{\hat{\psi}^{\dag}_L \hat{\psi}_R}$ cancel. Thus nonzero pion condensate can only develop at sufficiently strong coupling.

%\bibliographystyle{mybibstyle}
%\bibliography{Buividovich}

\end{document}